# Empowering Abilities: Increasing Representation of Students with Disabilities in the STEM Field

Esperanza Moreno[1], Piyush Kumar[1], Richard O Adansi[1], Dorothy Moreno[1], Demy Rodriguez[1], Raul Baez Ramirez[1], Audrey R Kapsa[1], Arturo Rodriguez[1], Neelam Agarwal[2], Vinod Kumar[3], Beverley A Calvo[1], and Vivek Tandon[1]

[1]The University of Texas at El Paso, Texas, 79968

[2]College of Alameda, Alameda, CA 94501

[3]Texas A&M University, Kingsville, Texas, 78363

## Abstract

The ExploreSTEM Summer Camps 2023 were designed to deliver inclusive STEM education to students aged 14 to 22 years with disabilities. This paper presents a thorough examination of the 2023 camp program, emphasizing the pivotal role of inclusive STEM education in potentially shaping students' personal and academic trajectories. The curriculum encompassed four weeklong fundamental STEM domains: Internet of Things (IoT), Computational Engineering, Artificial Intelligence (AI), and Augmented and Virtual Reality (AR/VR). Within Camp 1, students actively engaged with Dash robots, employing dedicated programming environments to command actions and gather sensor data, fostering interactions with the IoT platform and facilitating seamless data transmission. Camp 2 was dedicated to acquainting students with foundational computational engineering principles, establishing a robust framework for comprehending intricate engineering concepts. Camp 3 commenced with insightful presentations elucidating AI applications across multifaceted industries, including engineering, healthcare, and education, illuminating AI's pervasive influence on contemporary society. The primary aim of Camp 4 was to introduce students to the immersive domains of AR and VR, showcasing their applications beyond conventional STEM disciplines into everyday life experiences. The amalgamation of informative presentations, interactive activities, and a nurturing learning environment cultivated an engaging and enriching experience for all participants. By embracing inclusivity and harnessing innovative pedagogical approaches, the ExploreSTEM Summer Camps empowered students to explore, innovate, and excel within the dynamic realm of STEM education.

## 1. Introduction

In a rapidly evolving world driven by science, technology, engineering, and mathematics (STEM), every student deserves the opportunity to explore their potential, regardless of their abilities. Currently, people with disabilities are underrepresented in STEM occupations in comparison with their share of the general U.S. population, according to data from the National Science Foundation [1]. Addressing this underrepresentation in STEM fields has been an initiative of the U.S. Congress for the past 30 years, but the challenge remains unresolved. To address the low representation of students in STEM fields, intervention needs to happen before college. As our world becomes more technologically advanced, STEM is more important than ever and opens the door for individuals with disabilities to the most versatile and high-paying jobs available today [2]. Various strategies, including direct instruction, textbook-based education, and mnemonics, have been identified as less effective in closing the achievement gap, according to review studies [3,4]. With this vision in mind, a groundbreaking summer camp program has been established by authors of this paper, dedicated to empowering students with disabilities ages 14 to 22 through immersive STEM education. This innovative initiative aims to provide career-oriented guidance and experiential learning to high school age students with disabilities with the aim of opening doors to STEM related pathways s. By focusing on foundational concepts in cutting-edge fields such as the Internet of Things (IoT), Computational Engineering, AI in Engineering, Virtual Reality, and Augmented Reality, this program broadens horizons (opportunities) and creates pathways to success. The center of this summer program beats with a profound desire to illuminate the various educational and vocational avenues that STEM offers. Through the crafted modules, where participants explored concepts of Computational Engineering, where the language of algorithms and codes forms the basis of innovation, they. explored the integration of AI in Engineering, witnessing firsthand how machines can learn and adapt, thus harnessing their imagination to reshape the future. Virtual Reality and Augmented Reality transported them beyond the ordinary, while the Internet of Things (IoT) connected them to a web of interconnected devices that define modern existence.

At the core of this endeavor lies a belief in the power of interactive, hands-on learning [5]. The summer camp curriculum blended foundational theories with practical applications, ensuring that participants not only grasped abstract concepts but also engaged with technology in tangible ways. The mission of this transformative summer program extends far beyond the realm of academic enrichment [6]. It aspires to kindle a lifelong flame of interest in STEM, propelling participants towards future pursuits and careers they might not have once considered possible. By raising awareness of the diverse pathways within STEM [7,8], this initiative takes steps to dissolve the barriers that these students may aspire. Knowledge is the currency of empowerment [9], and this program is committed to not only imparting STEM skills but also nurturing confidence and self-belief. As participants gained insight, honed skills, and unlocked potential, they are poised to become ambassadors of change, proving that the limits of possibility are only defined by one's determination. Overall, the aim of our summer camps is to inspire participants to seek future STEM related pathways as they transition to post-secondary education.

ExploreSTEM summer camps were coordinated by The University of Texas El Paso (UTEP) research group, with support from the Texas Workforce Commission (TWC). The study reflects on the design of the weekly module and on the experiences of the program participants as they navigated the STEM topics, participated in everyday implementation, and integrated wellness and self-advocacy related topics. Challenges are further discussed which include initial communication barriers, hesitance in participation, and disruptive behavior. The team of leaders engaged in strategic approaches such as icebreakers, interactive polls, and curriculum adjustments to ensure daily challenges were addressed. Personalized support for individual learning needs of the students/participants were considered in the planning of activities and structuring the learning environment was crucial to maintaining student engagement and discipline. In the rapidly evolving landscape dominated by science, technology, engineering, and mathematics (STEM), providing equal opportunities for all students to realize their potential is paramount. To address the camp recruitment, the ExploreSTEM group recruited students at the TWC event and at some high schools that had the students with challenges program. The ExploreSTEM group conducted those activities before the camps started. If the students were interested, we had them follow up with the TWC to register with them first, and then parents would let us know what camps they would like to register their students for. Minimum attendance per week was 4 days out of 5, and every camp was weeklong.

This transformative summer program transcends academic enrichment. It aspires to ignite a lifelong passion for STEM, propelling participants towards previously unconsidered pursuits and careers. By promoting awareness of the diverse pathways within STEM, this initiative takes significant steps to break down barriers that may have once limited these students' aspirations. Nonetheless, knowledge is the currency of empowerment, and the program's culminating result and expectation is to impart basic STEM skills and nurture confidence and self-belief. As participants gain insight, refine their skills, and unlock their potential, they become ambassadors of change, demonstrating that the boundaries of possibility are defined only by one's determination.

This paper provides a comprehensive and chronological overview of the "Empowering Abilities" summer camp program, ExploreSTEM. Highlighting its transformative impact on students with challenges and its commitment to nurturing future engineers, regardless of their abilities.

## 2. Methodology

ExploreSTEM was established in response to the burgeoning interest and demand for STEM careers, particularly among students residing in El Paso, Texas, and border regions. This comprehensive initiative seeks to address the aspirations of these eager young minds, who, despite their immense potential, often encounter limitations in their pursuit of knowledge. At its core, ExploreSTEM is a multifaceted program designed to inspire and guide students toward promising STEM career pathways through a rich tapestry of educational experiences. Central to its mission is the creation of tailored programs and opportunities that empower students who may aspire to pursue future engineering post-secondary programs by immersing them in hands-on, problem-based STEM activities.

ExploreSTEM recognizes the intrinsic value of students' curiosity and their innate desire to explore the intricate world of STEM and engineering. It acknowledges that external circumstances may hinder these ambitions. Therefore, ExploreSTEM's approach is diverse and adaptable, intending to kindle enthusiasm for STEM among students through various channels. This initiative is deeply committed to broadening the horizons of students with disabilities, enhancing their STEM knowledge and skills, dismantling barriers, and instilling a lifelong passion for these fields.

The Texas Workforce Commission (TWC) provided financial support for the summer camps, and participation hinges on TWC registration. The meticulous planning, coordination, and execution of these camps are entrusted to an esteemed engineering research group (i.e., Advanced Modeling and Simulation Research Group) at The University of Texas at El Paso (UTEP). The camp activities are guided by Associate Professors, graduate, and undergraduate students from UTEP's engineering and education departments.

The preparation process for these summer camps is a well-choreographed endeavor, spanning approximately six months. A significant portion of this timeline is dedicated to fine-tuning and customizing activities to meet the specific needs of our diverse student cohort and other organizational duties such as scheduling and coordination of reservations, lunch reservations, extracurriculars, and more. Our curriculum unfolds structured, consisting of weekly camps running from Monday to Friday, commencing at 9 a.m. and concluding at 2 p.m., as thoughtfully outlined in the following tables.

| Table 1. ExploreSTEM Summer Camps 2023 | | Number of Students |
|---|---|---|
| Camp 1 | Internet of Things: June 12th - 16th, 2023 | 8 |
| Camp 2 | Computational Engineering: June 19th - 23rd, 2023 | 7 |
| Camp 3 | Artificial Intelligence in Engineering (AI): July 10th - 14th, 2023 | 7 |
| Camp 4 | Augmented Reality & Virtual Reality (AR/VR): July 17th - 21st, 2023 | 9 |

Within our curriculum, we carefully curated four essential STEM domains: Internet of Things (IoT), Computational Engineering, Artificial Intelligence (AI), and Augmented and Virtual Reality (AR/VR). The subject, and time of occurrence are described in Table 1. We employ a dual assessment approach to measure students' progress, encompassing pre- and post-assessments. Each week commences with administering a 10-question survey on Monday, serving as a baseline evaluation of students' comprehension of the topic. On Friday, marking the week's culmination, the same survey was revisited as a post-assessment, facilitating a measurement of knowledge acquisition.

Furthermore, our meticulously designed itineraries integrated diverse hands-on and computer-based activities, ensuring that the learning journey was dynamic and enriching. This approach enhances our coverage of STEM subjects and continuously elevates our instructional methods. We firmly believe that ongoing assessment of students' performance and knowledge is fundamental to delivering a superior educational experience. All questions featured in our assessments were thoughtfully aligned with the content explored throughout the camps, offering students an opportunity to review their knowledge and challenge their memory retention.

This paper extensively explores the instructional methods meticulously employed in our summer camps. These methods include interactive, hands-on activities, computer-based learning modules, and captivating presentations. We underscore the critical role of interactive learning in igniting students' enthusiasm, fostering a deep understanding of STEM careers and subjects, and nurturing essential skills in self-advocacy.

## 3. ExploreSTEM Summer camps

All our camps adhered to a structured schedule to ensure a seamless experience for ExploreSTEM staff and enthusiastic volunteers. The day commenced with their arrival between 8

a.m. and 9 a.m., during which staff set up the designated classrooms. This preparation included arranging materials and snacks in readiness and preparing presentations for the day's activities.

The camp activities started off at 9 a.m. and ran until 2 p.m., offering a comprehensive learning experience. Each day was planned, beginning with a focused session from 9 a.m. to 10 a.m. centered around a specific topic. These sessions featured engaging activities, which may have involved hands-on experiments or online exploration.

Following each hour-long session, a break to allow participants to stretch, enjoy a snack, hydrate, and converse with peers and instructors. To provide participants a full campus experience, lunches were provided in the UTEP Union buffet style cafeteria and lunch. Where they informally interacted with UTEP students. During lunch breaks, the students engaged in enjoyable interactions while also expressing their curiosity about the historical background of the campus, thereby contributing to a well-rounded and positive camp experience. Additional, on Fridays, participants were encouraged to suggest a special celebratory meal to recognize their weeklong achievements (e.g., ordering pizza, or walking to the campus Chick-Fill).

All participants received an ExploreSTEM t-shirt and personalized water cup. After completing four or more days of camp, participants received a participation incentive and a certificate related to the theme.

### 3.1 Camp 1: Internet of Things

The ExploreSTEM summer camps commenced with a focus on the Internet of Things (IoT) as the central theme for the week, as illustrated in Table 3, along with the detailed camp schedule. A pivotal aspect of our program was establishing an inclusive environment that welcomed staff members, volunteers, and participants. This endeavor posed notable challenges, marking the initial encounter for all involved. We initiated the program with a welcome ice-breaking presentation, offering an overview of what participants could anticipate throughout the camps. The program encompassed learning objectives, guided building and campus tours, dining arrangements, and incentives. Establishing a common ground for the students was deemed imperative. To foster a more open atmosphere, ExploreSTEM members introduced themselves, sharing their names, educational backgrounds, and intriguing facts about themselves. While some students remained initially reserved, others exhibited curiosity when we shared exciting tidbits, such as our passion for VEX robotics or video gaming. This insightful exchange allowed staff members to understand the students' interests, which were then incorporated into future camp activities.

| Table 2. Camp 1: Internet of Things (IoT) June 12-16, 2023 | | | | | |
|---|---|---|---|---|---|
| Date<br>Location | Monday, June 12<br>IDRB 2.215 | Tuesday, June 13<br>IDRB 2.215 | Wednesday, June 14<br>IDRB 2.215 | Thursday, June 15<br>IDRB 2.215 | Friday, June 16<br>IDRB 2.206 |
| Time | Activities | | | | |
| 8AM-9AM | Staff preparation (prep) for camp | Staff prep for camp | Staff prep for camp | Staff prep for camp | Staff prep for camp |
| 9AM-10AM | Welcome Presentation & Pre-Survey | STEM: Daily board & review of previous day | STEM: Daily board & review of previous day | STEM: Daily board & review of previous day | STEM: Daily board & review of previous day |
| 10AM-11AM | Non-STEM: Presentation | STEM: Intro to Programming | STEM: Intro to wonder workshops | Non-STEM Self advocacy presentation | STEM: Post-survey & Certificates |
| 11AM-12PM | Lunch | Lunch | Lunch | Lunch | Lunch |
| 12PM-1PM | STEM: Intro to IOT | STEM: Scratch basic terms & dash robots | STEM: Wonder app dash robots | STEM: Robot Competition part 1 | Dismissed |
| 1PM - 2PM | STEM: Explore Dash robots' activity & Wrap-up | STEM: create your own scratch code & Wrap-up | STEM: Blockly app dash robots | STEM: Robot Competition part 2 & Wrap-up | Dismissed |

Several participants required accommodations during the assessment phase, indicating diverse learning needs within the group. Some students were unable to respond independently to the assessments, highlighting the importance of personalized support. The accommodations requested included:

Extended Time: Some participants requested additional time, not only to complete assessments but also for activities like using the bathroom. This accommodation recognizes the diverse pacing of learners and ensures that everyone has sufficient time to engage with the material.

One-on-One Assistance: Certain students needed individualized support to understand assignments and complete STEM tasks. This accommodation acknowledges the varying levels of assistance required by different learners, promoting an inclusive learning environment.

Assistive Technology: Students with disabilities may require specific assistive technologies, such as screen readers, magnifiers, or specialized input devices, to access and engage with the assessment materials effectively.

Alternative Formats: Some participants may benefit from assessments presented in alternative formats, such as audio or tactile formats, to cater to diverse learning styles and accessibility needs.

Following the icebreakers and introductions, we proceeded to administer a pre-assessment. This assessment consisted of questions about the Internet of Things to gauge the students' baseline knowledge. We deliberately refrained from offering immediate answers to students' questions, instead encouraging them to respond to the best of their knowledge. This approach ensured that we began with an appropriate pace for introducing IoT concepts. This phase encompassed approximately one hour before transitioning to the subsequent section, where we delved into the practical aspects and challenges associated with IoT, offering potential solutions.

For the first week of camp, eight students participated. To assess the students' baseline knowledge of IoT, we administered a pre-survey at the outset. This survey played a pivotal role in gauging their initial understanding. It paved the way for a post-survey, which we conducted to measure the progress and learning outcomes achieved during the camp. Throughout the week, the students were given an informative presentation covering various facets of IoT, including its engineering applications, everyday uses, historical context, and significance in domains such as education and healthcare (see Figure 1). The dynamic presentation style effectively captivated the students' attention and facilitated a more profound comprehension of STEM subjects.

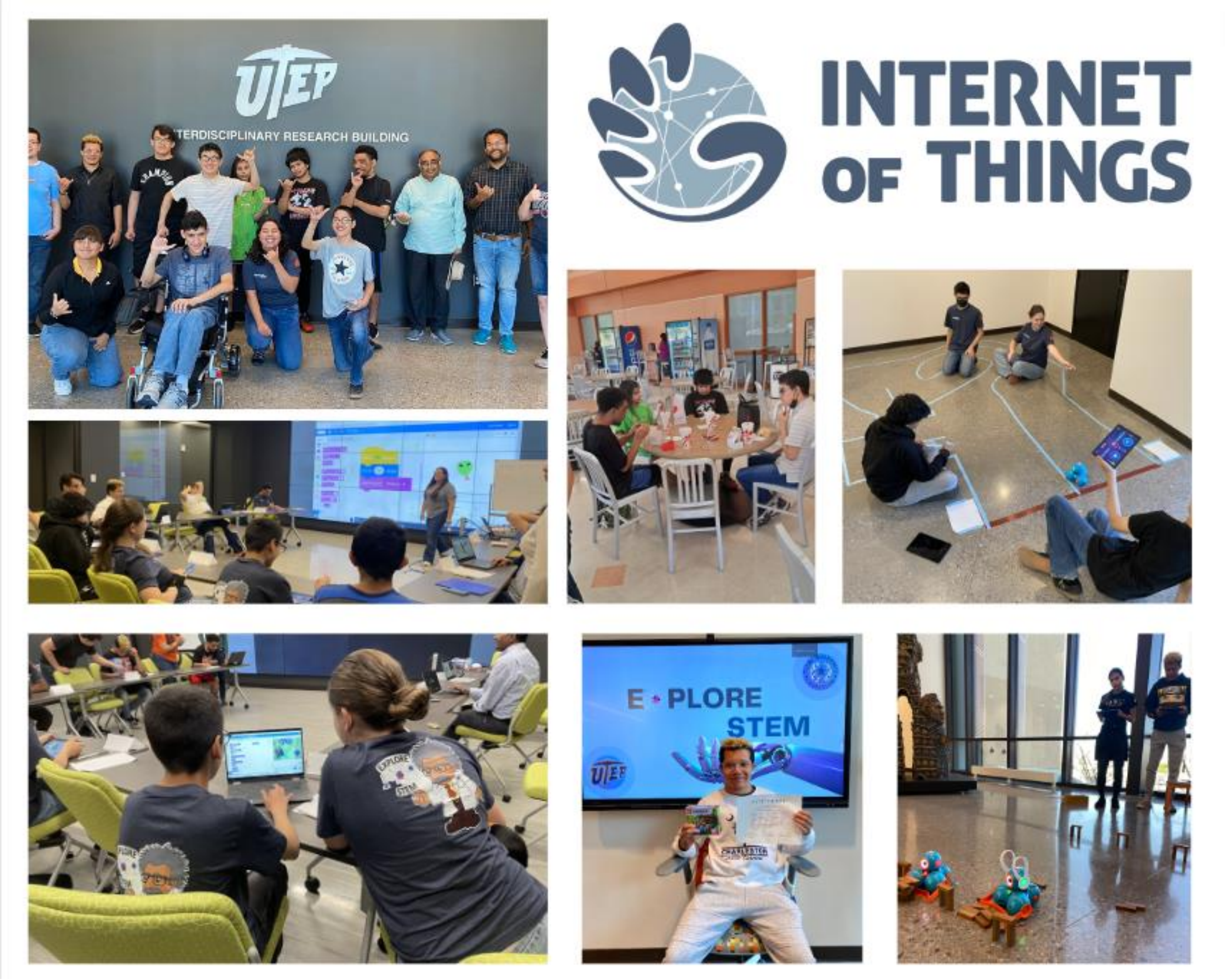

**Figure 1.** Pictures of the activities, lunch, and awards from Camp 1: Internet of Things

A particular highlight of the week was emphasizing the practical application of concepts using Dash robots. These robots served as valuable tools for the students to create lines, shapes, and practical scenarios, enabling them to gain hands-on experience and hone their skills. Additionally, the Dash robots were equipped with programming environments, empowering the students to control the robots' actions and receive sensor data, facilitating interactions with the IoT platform, and enabling data transmission. The introduction of coding and robotics concepts was conducted dynamically and engagingly, further fueling the students' enthusiasm and inquisitiveness. Their palpable interest in exploring the Dash robots' possibilities underscored this activity's success. 80% student were successful in programming the Dash robot. In alignment with the core content, our educational resources incorporated essential vocabulary, facilitating the students' comprehension of the subject matter. Key terms such as cloud servers, internet sensors, connectivity, data processing, and user interfaces were thoughtfully introduced, contributing to a more comprehensive understanding of the intricacies of the Internet of Things.

The overall camp experience was a resounding success, as evidenced by the students' gradual transition from initial reserve to active engagement and peers interaction. The combination of informative presentations, practical activities, and the nurturing of a welcoming atmosphere contributed to a highly productive and enjoyable learning environment for all participants, including those with challenges.

In the context of non-STEM activities, during the first week of ExploreSTEM, students participated in a session centered on "Students with Challenges: Barriers & Opportunities in STEM." This presentation included the definitions of challenges, challenges encountered in high school and post-secondary education, and potential resources and support mechanisms for students with challenges, such as academic programs, CAAST, and scholarships.

Therefore, the presentation encompassed an examination of the challenges and prospects faced by students with challenges within the fields of Science, Technology, Engineering, and Mathematics (STEM). The topic delved into the obstacles and difficulties these students encounter when pursuing education and careers in STEM, while also exploring the potential avenues and positive aspects that can empower them to excel in these disciplines. We encouraged a thorough exploration of the difficulties impeding their progress and the possible breakthroughs and inclusive strategies that can create a more equitable and supportive environment for their involvement in STEM fields.

### 3.2 Camp 2: Computational Engineering

The Computational Engineering camp's objective was to familiarize students with the concept of computational engineering within the framework of STEM education. Computational thinking, a fundamental aspect of this endeavor, involves comprehending how computers and software systems process information by deconstructing intricate problems into manageable steps called algorithms. Due to the federal holiday this was a 4-day camp session. To foster an inclusive and neutral environment, we acknowledged the significance of June 19th as a holiday for our students, making it the second week of ExploreSTEM. Seven students participated this week (see Table 2), five returning from the previous camp.

| Table 3. Camp 2: Computational Engineering June 19-23, 2023 | | | | | |
|---|---|---|---|---|---|
| Date Location | June 19th Holiday | 20 - IDRB 2.215 | 21 - IDRB 2.215 | 22 - IDRB 3rd floor | 23 - IDRB 2.206 |
| Time | Activities | | | | |
| 8AM-9AM | No Class | Staff prep for camp | Staff prep for camp | Staff prep for camp | Staff prep for camp |
| 9AM-10AM | No Class | Welcome Presentation & Pre-Survey | Non-STEM: Pawsitive Connections | Non-STEM: S.Y.T.Y.K.College | Non-STEM: college exploration |
| 10AM-11AM | No Class | STEM: Intro to Comp Engineering | STEM: Intro to CAD | Non-STEM: Kahoot Knowledge Skit | Non-STEM: Mark Jurado presentation |
| 11AM-12PM | No Class | Lunch | Lunch | Lunch | Lunch |
| 12PM-1PM | No Class | STEM: Intro to Algorithms | STEM: Sculpting & Modeling activity | STEM: Intro to Tinkercad | STEM: Post-survey & Certificates |
| 1PM - 2PM | No Class | STEM: Cups in Algorithm activity | STEM: Nomads CAD & vocabulary | STEM: Creations on Tinkercad activity | STEM: GAIA Tour activity |

Throughout this week, students actively participated in both hands-on activities and computer-based learning to delve into the above-mentioned concepts. A standout activity during the week was "STEM Coding for Students - Cup Stacking Algorithms," which involved creating cup stacking designs and formulating algorithms to construct them. This activity reinforced their grasp of computational thinking and promoted teamwork and collaboration among the students. Additionally, the camp featured activities such as "Building Reflections," during which students discussed the challenges they encountered while developing their algorithms. This environment facilitated supportive interactions and the sharing of experiences, promoting social engagement alongside STEM education. The daily routine had become familiar for returning students, with a whiteboard student engagement activity involving daily polls related to the weekly main topic. We also conducted other presentations, such as one on video gaming, which covered the broader concept of it, its societal importance, safe gameplay practices, and potential careers in the field. While centered around video games, this topic resonated with our students and complemented computational engineering discussions. Several undergraduate computer science majors shared their academic journeys, inspiring our students to follow similar paths.

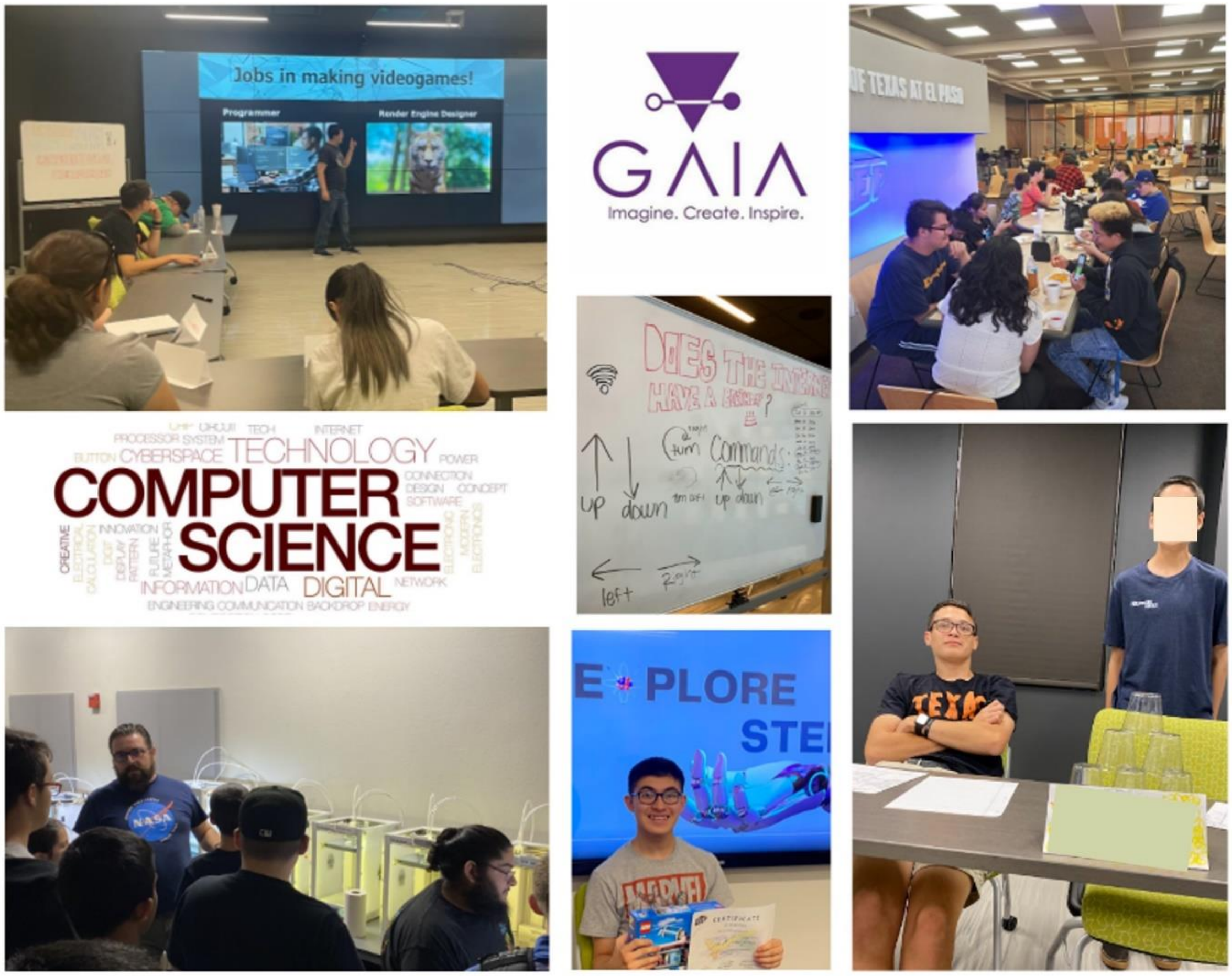

**Figure 2.** Pictures of the activities, lunch, and rewards from Camp 2: Computational Engineering

To provide students with a broader perspective, we explored how Computer-Aided Design (CAD) and 3D printing technologies are utilized in creating three-dimensional objects. ThinkerCAD, a user-friendly CAD software, played a pivotal role in this process. This hands-on approach allowed students to witness the transformation of their virtual designs into physical objects. To enhance the learning experience, we organized a tour of GAIA Services at the University of Texas at El Paso (UTEP). During this tour, students were exposed to various advanced resources, including 3D printing, laser cutting, bandit gaming teams, engraving, and more. Witnessing the extensive range of tools available at Gaia Services amazed the students. As a culmination of their learning, students had the opportunity to 3D print their ThinkerCAD creations at Gaia Services, leaving them both educated and thrilled. The student behavior transformation was evident, transitioning from timidity to open enthusiasm for peers collaboration. Exploring different areas of the campus instilled in them the aspiration to return as not just visitors but as future students. Figure 2 shows the pictures of the activities, lunch, and awards from Camp 2: Computational Engineering.

In addition to STEM-focused activities, we provided non-STEM sessions that covered the distinctions between high school and college and essential college preparation tips. This allowed students to organize their thoughts about their future education, and we offered guidance and assistance in addressing any queries related to enhancing their college readiness. Furthermore, our visit to the technology center introduced students to a technology professor and offered insights through a college major assessment, providing an enjoyable and empowering experience. The program also included an informative segment about service dogs, with the participation of two guests and their service dogs. We highlighted these support animals' diverse sizes and roles, emphasizing the assistance they can provide to students in their current or future educational journeys. Lastly, a guest speaker shared her experiences at UTEP's Center for Accommodations and Student Services (CASS), where accommodation and student support requests are processed. This provided valuable information on the support systems available to students, ensuring they have access to the necessary resources. In conclusion, our comprehensive approach guided the students through the intricacies of computational engineering, transforming their behavior and aspirations. This holistic educational experience combined STEM education with invaluable insights into college preparation and support services, equipping our students for a brighter future.

### 3.3 Camp 3: Artificial Intelligence in Engineering (AI)

The Artificial Intelligence in Engineering (AI) camp was designed to offer students a comprehensive initiation into the captivating realm of Artificial Intelligence (AI). Our primary objective was to showcase AI's multifaceted capabilities and empower students with the knowledge necessary to access and harness this transformative technology. Through engaging activities (see Table 5), we aimed to equip participants with a solid foundational understanding of AI and its myriad applications.

| Table 4. Camp 3: Artificial Intelligence in Engineering (AI) July 10-14, 2023 | | | | | |
|---|---|---|---|---|---|
| Date Location | 10 - IDRB 2.215 | 11 - IDRB 2.215 | 12 - IDRB 2.215 | 13 - IDRB 2.206 | 14 - IDRB 2.206 |
| Time | Activities | | | | |
| 8AM-9AM | Staff prep for camp | Staff prep for camp | Staff prep for camp | Staff prep for camp | Staff prep for camp |
| 9AM-10AM | Non-STEM: UTEP CASST | Non-STEM: Project Amistad Presentation | Non-STEM: MRC Presentation | Non-STEM: Budgeting with amistad project | STEM: Eric presentation |
| 10AM-11AM | Welcome Presentation & Pre-Survey STEM: Intro to AI | STEM: AI in healthcare field | STEM: Robotic hand basic anatomy and hands on activity | STEM: Synthesia & Steve AI exploration | STEM: Post Assessments & Wrap up |
| 11AM-12PM | Lunch | Lunch | Lunch | Lunch | Lunch & chat with parents |
| 12PM-1PM | STEM: Quick draw web apps in AI | STEM: Intro to robotic hand project | STEM: wrap up robotic hand & Intro to AI video generator | STEM: Gratitude wall and dialog creation for text to speech video | Lunch & chat with parents, certificate distribution |
| 1PM - 2PM | Non-STEM: Elevator speech with ChatGPT | Non-STEM: Elevator Speech ChatGPT | Non - STEM: Self-Motivation | Non - STEM: Eric Alonzo Presentation | Presentations for parents & AI video creations |

The camp commenced with informative presentations on various AI applications across engineering, healthcare, education, and daily life domains, demonstrating how AI shapes industries and influences the world. Students were exposed to compelling videos illustrating AI's pivotal role in developing robots and prosthetic arms, highlighting its potential to revolutionize human-machine interaction. To ensure a robust comprehension of the subject, we introduced fundamental AI terminology, laying the groundwork for students' future AI-related projects and endeavors. One of the camp's most engaging activities was the "Robotic Hand Project." This project allowed participants to explore the intersection of robotic technology and AI while also examining how these advancements can enhance the lives of individuals with challenges. Guided by STEM instructors, students began constructing their robotic hands using paper and assorted materials. This hands-on exercise fostered an understanding of the mechanics involved and sparked curiosity about AI's role in modern innovations. The "Steve AI" activity centered on the creative utilization of AI for video production in the context of STEM. Through this endeavor, participants discovered how machine learning can be harnessed to create videos that reflect their interests and passions. Of particular significance was this activity's positive impact on students with challenges. "Steve AI" provided a novel platform for self-expression and learning, allowing them to generate videos from simple text scripts. The students embraced this innovation vis-a-vis their unwillingness to present in front of an audience but were eager to share their enthusiasm for STEM and other hobbies.

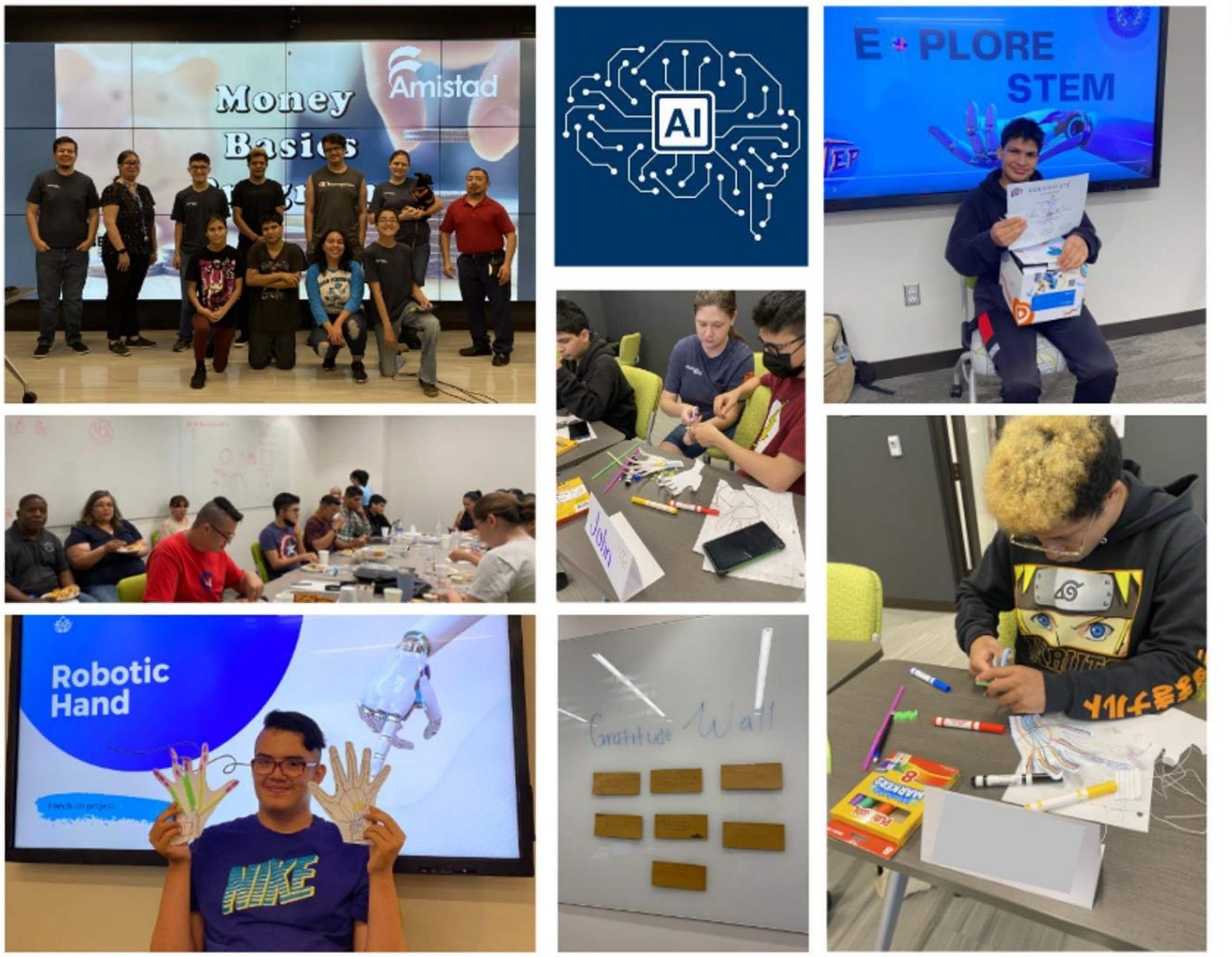

**Figure 3.** Pictures of the activities, lunch, and rewards from Camp 3: Artificial Intelligence in

Engineering (AI).

Outcomes of the AI camp included a dynamic fusion of informative presentations and engaging activities. Students exhibited heightened interest in both the mechanics of the robotic hand project and the intricate involvement of AI (see Figure 3). The "Steve AI" activity empowered students to explore creative expression and enabled them to generate AI-driven videos showcasing their personalities and passions. Furthermore, the camp extended its impact, as students were eager to apply the skills acquired in their projects and other pursuits. Parents also gained an exclusive glimpse into the camp's activities through the AI-generated videos, offering them insights and perspectives on their children's enriching experiences.

In conclusion, the camp successfully achieved its goal of introducing students to the captivating world of Artificial Intelligence (see Figure 3). By blending informative sessions with hands-on activities, we fostered a robust foundational understanding of AI and empowered students to explore its creative applications. The camp ignited curiosity about AI's potential and provided a platform for inclusive learning and expression. As we reflect on the camp's accomplishments, our hope is that the knowledge gained, and experiences shared will continue to inspire these young minds on their journey into the ever-evolving realm of AI.

Unfortunately, an issue arose with one student who required additional attention due to their condition. While most students gradually became comfortable with the teachers and adhered to the rules, this particular student's behavior became disruptive. Instead of guiding him back on track with the lesson, some helpers allowed him unrestricted internet access, leading to disruptions in the class and causing concerns among his classmates. As a summer program focused on increasing awareness of STEM opportunities for students with disabilities, it is essential to maintain a level of discipline and respect in the learning environment. We designed pre- and post-assessments to evaluate students' progress and understanding of STEM topics. However, the mentioned student refused to comply, citing non-STEM activities to avoid assessments. Establishing a solid behavioral expectations management system for all students, including those with challenges, is crucial to addressing such challenges and creating an effective educational system. A structured approach to behavior management, tailored to the unique needs of students with challenges, can foster a positive learning environment. Proper training and communication with teachers, helpers, and students can promote understanding, empathy, and mutual respect. Implementing consistent boundaries and consequences can prevent students from exploiting their challenges as excuses for inappropriate behavior.

In conclusion, the camp successfully introduced AI concepts to students with challenges and highlighted the relevance of AI in various fields. However, it also underscored the importance of maintaining discipline and respectful behavior within the educational setting, particularly for students with different needs. By establishing a well-designed education system with appropriate behavioral guidelines, educators can enhance the learning experience for all students and create an inclusive and supportive STEM environment.

Non-STEM: During the camp, we actively participated in activities guided by our guest from Project Amistad, aiding students in understanding the steps to improve their financial management skills. Additionally, one of our volunteers delivered a presentation on Rehabilitation Counseling, leveraging her expertise in this area. In collaboration with the STEM group, we undertook a meaningful task: helping students craft elevator pitches using ChatGPT to refine their ideas. The subjects covered diverse areas, including money management, life skills, and strategies for achieving financial independence. The interactive approach captivated the participants, eliciting their enthusiasm.

### 3.4 Camp 4: Augmented Reality & Virtual Reality (AR/VR)

The Augmented and Virtual Reality (AR/VR) camp's primary goal was to introduce students, including those with challenges, to the dynamic realms of Augmented Reality (AR) and Virtual Reality (VR), with a strong emphasis on STEM education. These innovative technologies hold diverse applications, extending beyond STEM disciplines into everyday life. For students with challenges, AR and VR offer unique advantages in addressing academic challenges and fostering self-advocacy skills, rendering the camp an inclusive and captivating experience for all participants. Table 4 shows a detailed description of Camp 4: Augmented Reality & Virtual Reality.

The camp leveraged Oculus Go headsets to facilitate this immersive journey, providing students with an exceptional opportunity to immerse themselves in a diverse range of VR experiences. These experiences encompassed captivating encounters with prehistoric creatures and immersive documentaries showcasing the wonders of the natural world. For students with challenges, VR presented an alternative and more accessible avenue for exploring and engaging with complex educational content, thereby promoting equity within the classroom. While students enjoyed the VR technology, they also expressed a preference for hands-on activities. To address this, we introduced activities such as the merging AR cube and coloring VR, which offered a more relaxed and artistic approach to AR/VR, allowing students to witness their creations come to life on mobile devices.

| **Table 5. Camp 4: Augmented Reality & Virtual Reality (AR/VR) July 17 - 21, 2023** | | | | | |
|---|---|---|---|---|---|
| Date Location | 10 - IDRB 2.215 | 11 - IDRB 2.215 | 12 - IDRB 2.215 | 13 - IDRB 2.206 | 14 - IDRB 2.206 |
| Time | Activities | | | | |
| 8AM-9AM | Staff prep for camp | Staff prep for camp | Staff prep for camp | Staff prep for camp | Staff prep for camp |
| 9AM-10AM | Non-STEM: Deciphering the ADA | Non-STEM: My best college fit with ADA | Non-STEM: O*Net Presentation | Non-STEM: Career Exploration | Non-STEM: Self Worth |
| 10AM-11AM | STEM: Intro to AR/VR | STEM: Intro to AR in Pokémon go | STEM: Intro to VR headsets | STEM: Intro to Federick Lanchester & VR tour | Non-STEM: Dr. Calvo talk |
| 11AM-12PM | Lunch | Lunch | Lunch | Lunch | Lunch |
| 12PM-1PM | STEM: AR vs VR & quiver vision | STEM: Pokémon Go & safety training | STEM: VR headset on Jurassic Park | STEM: Federick Lanchester hands on activity | STEM: Paper airplanes FL competition |
| 1PM - 2PM | Non-STEM: ADA Kahoot | Non-STEM: Animal & Crayon activity | Non-STEM: O*NET Assessment | STEM: VR headset individual exploration | STEM: Merge cubes, post-survey & Certificates |

Concurrently, the camp devoted time to an in-depth exploration of Augmented Reality, utilizing the popular "Pokémon Go" phenomenon as an educational tool (see Figure 4). Augmented Reality has the potential to bolster self-advocacy skills for students with challenges by providing real-world, context-based information. This technology aids them in navigating unfamiliar environments and gaining a better understanding of their surroundings, thereby empowering them to advocate for their needs effectively. Integral to this exploration was an examination of the prerequisites for pursuing degrees in engineering, particularly in game development. This discussion was supplemented by a thoughtful examination of the potential hazards associated with excessive gaming, including its impact on students with challenges. Understanding these nuances is essential for educators and caregivers to provide tailored support to students with challenges.

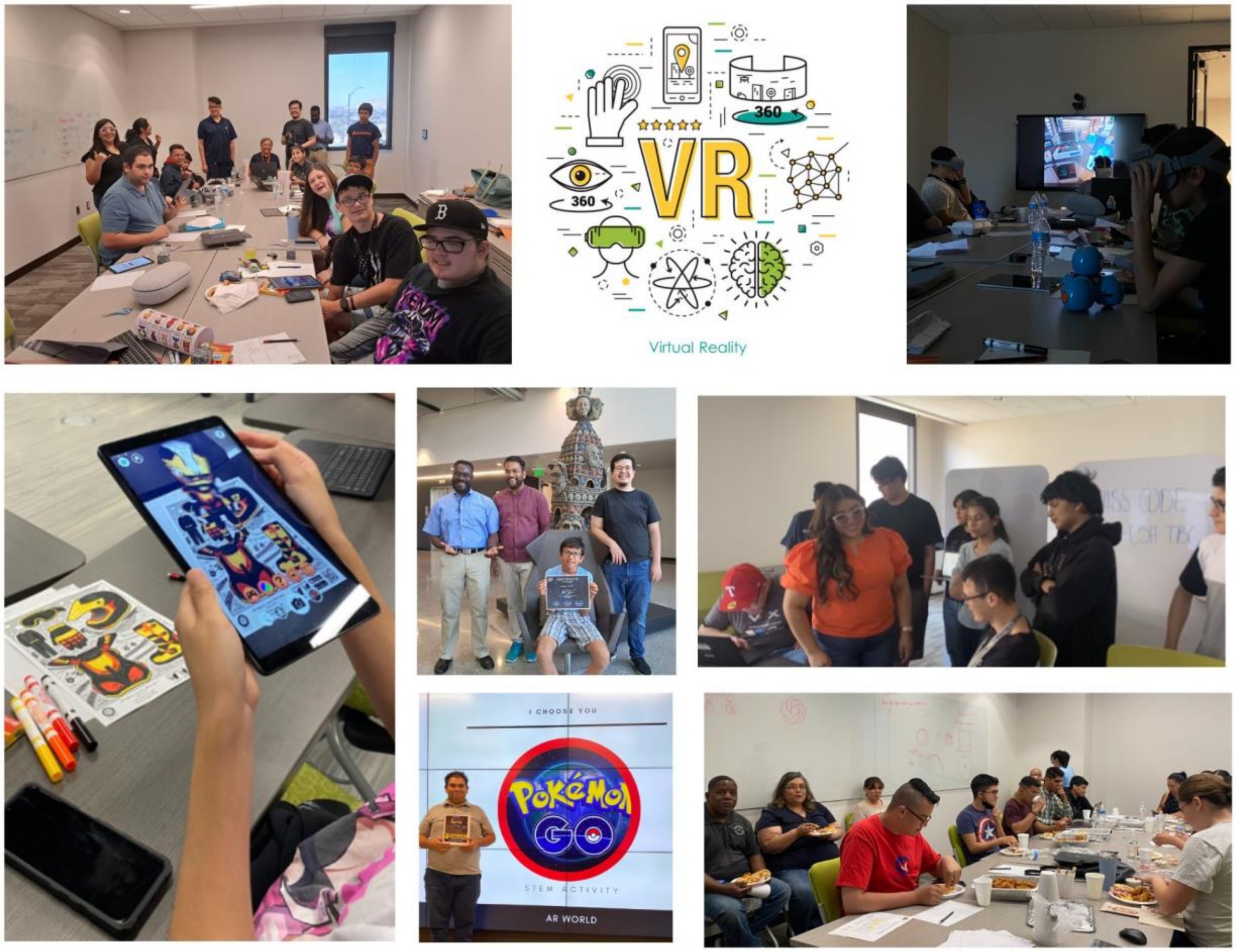

**Figure 4:** Pictures of the activities, lunch, and rewards from Camp 4:  Augmented Reality & Virtual Reality (AR/VR).

Moreover, the camp featured an immersive activity centered around the influential contributions of Frederick Lanchester, whose theories continue to resonate in the AR/VR domain. This activity utilized the virtual interface of Oculus Go to introduce students, including those with challenges, to Lanchester's concepts. It allowed them to apply these principles practically by designing and constructing paper airplanes (see Figure 4). This hands-on endeavor fostered creative thinking and provided insights into the intricate forces governing aircraft during flight, promoting problem-solving skills that are invaluable for all students, including those with challenges.

In summary, the camp successfully unveiled the potential of AR and VR to students, including those with challenges, igniting their curiosity and deepening their understanding. Simultaneously, it underscored the vast possibilities of these technologies within STEM but also in broader educational contexts. For students with challenges, AR and VR are potent tools for surmounting academic challenges and enhancing self-advocacy skills. The transformative impact of AR and VR on education, by enhancing engagement, interactivity, and overall effectiveness, is particularly noteworthy for students with challenges. Emphatically, the employment of AR and VR underscores the significance of continued exploration and integration of these technologies into inclusive educational settings.

## 4. Results

The ExploreSTEM Summer Camps 2023 were established to provide inclusive STEM education to students aged 14 to 22, with a focus on addressing academic challenges through immersive learning experiences across key STEM domains: Internet of Things (IoT), Computational Engineering, Artificial Intelligence (AI), and Augmented and Virtual Reality (AR/VR). This section presents the results of the program, aiming to provide a comprehensive overview of the outcomes achieved while addressing the challenges encountered.

At the outset, it's crucial to restate the problem that drove the purpose of the ExploreSTEM program: the underrepresentation of individuals with disabilities in STEM occupations despite their potential and the importance of providing equal opportunities for all students to explore their potential in STEM fields. The camps achieved significant milestones, demonstrating their profound impact on participants, particularly those facing challenges. Participants showcased remarkable growth across various aspects:

Through purposeful icebreakers and interactive polls, there was a notable enhancement in communication, fostering a more inclusive learning environment and bridging knowledge gaps. This improvement facilitated engaging discussions centered on STEM themes, promoting student interaction. Moreover, students exhibited heightened curiosity and interest in STEM topics, showcasing increased engagement. Their eagerness to attend future camps independently reflected elevated motivation and enthusiasm for learning. The program also fostered the formation of meaningful friendships, creating a supportive and inclusive community where participants showcased self-reliance and academic confidence. Emphasizing teamwork and support enriched the learning experience, promoting student collaboration and mutual support. Participants approached learning enthusiastically and reflected a positive attitude toward education throughout the program. Despite challenges, strategic approaches such as purposeful icebreakers, interactive polls, and curriculum adjustments effectively addressed communication barriers and hesitance in participation.

Personalized support was provided to students with unique conditions, although challenges such as disruptive behavior necessitated disciplinary measures. Assessments introduced to gauge comprehension demonstrated participants' academic progress and enthusiasm for STEM subjects. Visual representations such as graphs showing communication improvement over time or photos capturing collaborative activities and student engagement can be included to illustrate the findings further. In conclusion, the ExploreSTEM Summer Camps 2023 effectively promoted communication, independence, collaboration, and enthusiasm among participants, particularly those facing challenges. The program fostered holistic growth and development by embracing inclusivity and innovative pedagogical approaches, emphasizing the importance of tailored strategies for inclusive STEM education.

Overall feedback from student and their parents:

- Student X1: I enjoy the camp thank you for all of attention and help to find a stem place.
- Student X2: I love robots and thank you for the robot incentive.
- Student X3: I enjoy the cup bulging activities and also the robot competition.
- Student X4: We should have more video games and tablets.
- Student X5: I liked the camp just more individual time.
- Student X6: Thank you it was a great camp.
- Mother of Student X1: I saw a big change in X1 dealing with different teachers. He considered you guys teachers, so he respected you guys, So I think that's also for him to understand. I'm happy. And he really loved it. He was scheduled to go out of town, and he didn't go because Student X1 told me : , mama, I'm not going out of town I will go to the camps .
- Mother of Student X3: It was amazing. He was excited. He loved it. He wanted to keep coming back.

## 5. Conclusion

In summation, the ExploreSTEM Summer Camps of 2023 have yielded positive outcomes, serving as a testament to the potency of tailored STEM education programs that cater to a diverse range of students, including those with challenges. These camps, spanning four distinct domains – Internet of Things (IoT), Computational Engineering, Artificial Intelligence (AI), and Augmented and Virtual Reality (AR/VR), have unequivocally demonstrated their capability to provide enriching and inclusive STEM education. The pedagogical methods employed throughout these camps have elucidated the critical importance of interactive learning. By offering diverse instructional tools, from hands-on activities to computer-based learning modules and captivating presentations, the camps have succeeded in igniting students' enthusiasm and imparting a profound understanding of STEM subjects, complemented by nurturing essential self-advocacy skills.

Throughout these four camps, students have achieved exceptional milestones indicative of the program's effectiveness. These milestones encompass a broad spectrum of achievements, ranging from the notable enhancement of communication skills and the heightened interest in STEM topics to the formation of meaningful friendships and the inculcation of self-sufficiency in their academic pursuits. These accomplishments bear testament to the efficacy of the program's strategies in surmounting challenges and empowering students. However, it is essential to acknowledge the challenges encountered during these camps. These challenges underscore the necessity of maintaining discipline and setting clear behavior expectations applicable to all students, irrespective of their challenges.

In a world where STEM disciplines continue to evolve and define the future, the ExploreSTEM Summer Camps of 2023 have provided students with disabilities with the foundational tools, knowledge, and motivation to thrive and excel in these dynamic fields. These camps have functioned as educational initiatives and beacons of hope and empowerment for students facing various challenges, emphasizing that participants in this program can carve out a successful STEM path with adequate support and tailored strategies. As we reflect on the myriad achievements and exceptional milestones reached during these camps, it is clear that inclusive STEM education is an ideal and a palpable reality that empowers students to envision a future illuminated by the possibilities of science, technology, engineering, and mathematics. The triumphant saga of ExploreSTEM Summer Camps stands as a compelling model for future endeavors in inclusive STEM education, highlighting the transformative potential of such programs in nurturing the next generation of innovators, problem solvers, and leaders in the STEM fields. With resounding success as its hallmark, ExploreSTEM paves the way forward, reminding us that the pursuit of knowledge knows no bounds, and with dedication and adaptability, every student can shine in the realm of STEM. Therefore, it is imperative for the STEM community to embrace and expand such programs, recognizing their significance in addressing talent and capability gaps and furthering research in inclusive STEM education.

**Acknowledgment**

We acknowledge the Texas Workforce Commission, U.S. Department of Defense (AFOSR Grant Number # FA9550-22-1-0018, FA9550-19-1-0304, FA9550-17-1-0253, FA9550-17-1-0393 FA9550-12-1-0242 SFFP, AFTC, HAFB/HSTT, AFRL, HPCMP), U.S. Department of Energy (GRANT13584020, DE-SC0022957, DE-FE0026220, DE-FE0002407, NETL, Sandia, ORNL, NREL), Systems Plus, and several other individuals at these agencies for partially supporting our research financially or through mentorship. We would also like to thank NSF ((HRD-1139929, XSEDE Award Number ACI-1053575), TACC, DOE, DOD, HPCMP, University of Texas STAR program, UTEP (Research Cloud, Department of Mechanical Engineering, Graduate School & College of Engineering) for generously providing financial support or computational resources. Without their generous support, completing the milestones would have been almost impossible.